\documentclass[a4paper,11pt]{article}

\usepackage[T1]{fontenc}
\usepackage[utf8]{inputenc}
\usepackage{lmodern}

\usepackage{geometry}
\usepackage{booktabs}
\usepackage{array}
\usepackage{tabularx}

\usepackage{graphicx}
\usepackage{amsmath}
\usepackage{amssymb}
\usepackage{url}
\usepackage[numbers,sort&compress]{natbib}
\usepackage{lineno}
\usepackage[hidelinks]{hyperref}

\usepackage{tikz}
\usetikzlibrary{positioning, arrows.meta, calc, shapes.geometric}

\newcolumntype{L}[1]{>{\raggedright\arraybackslash}p{#1}}

\begin{document}

\title{OC-AND: A Citation-Enriched Benchmark for Author Name Disambiguation}

\author{
Francesca Cappelli$^{1}$ \quad Giovanni Colavizza$^{1,2,3}$ \quad Silvio Peroni$^{1,3,4}$
}

\date{%
$^1$Department of Classical Philology and Italian Studies, University of Bologna, Via Zamboni 32, 40126 Bologna, Italy\\[4pt]
$^2$Department of Communication, University of Copenhagen, Karen Blixens Plads 8, 2300 Copenhagen, Denmark\\[4pt]
$^3$Digital Humanities Advanced Research Center (DHARC), University of Bologna, Via Zamboni 32, 40126 Bologna, Italy\\[4pt]
$^4$Research Centre for Open Scholarly Metadata,Department of Classical Philology and Italian Studies, University of Bologna, Via Zamboni 32, 40126 Bologna, Italy\\[4pt]
}

\maketitle

\begin{abstract}
Author Name Disambiguation benchmarks are predominantly derived from closed infrastructures that may not reflect conditions in open bibliographic environments. We present OC-AND, an open citation-enriched benchmark for AND constructed by remapping the WhoIsWho dataset with bibliographic and citation metadata from OpenCitations. The dataset is produced through a reproducible five-stage pipeline involving DOI verification, metadata validation, OpenCitations enrichment, and author-centric transformation. OC-AND retains the WhoIsWho ground-truth files and partition structure unchanged, while its publication files are restricted to publications that could be matched to a validated DOI and an OpenCitations record. It introduces realistic characteristics of open environments: heterogeneous metadata completeness, absent abstracts and affiliations, asymmetric citation coverage, and sparse graph connectivity. The dataset contains author records with associated citations, enabling evaluation of AND methods in settings closer to real-world open scholarly infrastructure scenarios. OC-AND is released as open data with complete provenance documentation and pipeline code, supporting reproducible research and the development of citation-aware disambiguation approaches.
\end{abstract}

\section*{Background \& Summary}
Author Name Disambiguation (AND) is a fundamental task in digital libraries, scholarly knowledge graphs, and bibliometric infrastructures  ~\cite{torvik2009, strotmann2012,abramo2023reliable}. Its objective is to partition publications associated with ambiguous author name strings into distinct clusters corresponding to real-world individuals. Errors in this process propagate silently across downstream systems, affecting citation analysis, author profiling, collaboration networks, and research assessment ~\cite{strotmann2012, kim2018evaluating}.

Over the last decade, progress in AND has been driven by a small number of widely adopted benchmarks mostly derived from the AMiner infrastructure (\url{https://www.aminer.org/})~\cite{zhang2018}. As discussed in recent surveys~\cite{cappelli2025survey, ferreira2012brief, debonis2023graph}, these benchmarks have become the de facto evaluation standard for modern disambiguation systems. However, they suffer from systematic limitations that compromise their suitability for citation-aware and graph-based evaluation.

First, legacy datasets are tightly coupled to closed or partially accessible metadata ecosystems, limiting interoperability with open scholarly infrastructures. Second, citation information is often unavailable or weakly integrated, despite its growing importance in graph-based AND models ~\cite{debonis2023graph, schulz2014exploiting}. Third, metadata fields in AMiner-derived benchmarks are typically dense and consistently populated, producing artificially clean relational environments that differ substantially from the heterogeneous conditions observed in real-world open bibliographic data.

A parallel effort toward more interoperable evaluation environments is represented by the S2AND framework~\cite{subramanian2021s2and}, which unified historical datasets using the dense proprietary metadata layer of Semantic Scholar. However, such platforms operate in environments where metadata completeness is systematically managed by automated internal pipelines, leaving a methodological gap in understanding how disambiguation models behave when evaluated directly on real-world bibliographic settings~\cite{sanyal2019review}.

We introduce the OC-AND dataset, which systematically re-maps the entity architecture of WhoIsWho benchmark na-v3 Name Disambiguation from Scratch schema~\cite{chen2023whoiswho} onto OpenCitations (\url{https://opencitations.net/})~\cite{peroni2020opencitations}, an independent not-for-profit infrastructure providing open bibliographic and citation metadata through interoperable Semantic Web technologies, combining citation indexes and bibliographic metadata (OpenCitations Meta~\cite{massari2023opencitations}). Rather than carrying over the native AMiner metadata, our approach completely reconstructs the operational feature layer using exclusively OpenCitations bibliographic and citation records. Crucially, the OC-AND dataset directly inherits the ground-truth author clusters and partition assignments (training/validation split) from the original WhoIsWho benchmark na-v3. The result is a benchmark that intentionally retains metadata incompleteness, absent abstract and affiliation data, asymmetric citation coverage, and structural consequent graph sparsity as authentic characteristics of open bibliographic infrastructures.

The remainder of this paper is organised as follows. The \hyperref[sec:methods]{Methods} section describes the five-stage pipeline used to construct OC-AND, from gold-standard DOI recovery to the author-centric transformation. The \hyperref[sec:data_records]{Data Records} section details the structure of the released files, and the \hyperref[sec:dataset_profile]{Dataset Profile and Properties} section characterises metadata completeness and compares OC-AND with WhoIsWho across partitions. The \hyperref[sec:technical_validation]{Technical Validation} section assesses the accuracy of DOI assignment and a functional validation, while the \hyperref[sec:usage_notes]{Usage Notes} section describes the evaluation framework and reports baseline experiments. Finally, the \hyperref[sec:conclusions]{Conclusions} summarise the contribution and outline future work.

\section*{Methods}
\label{sec:methods}

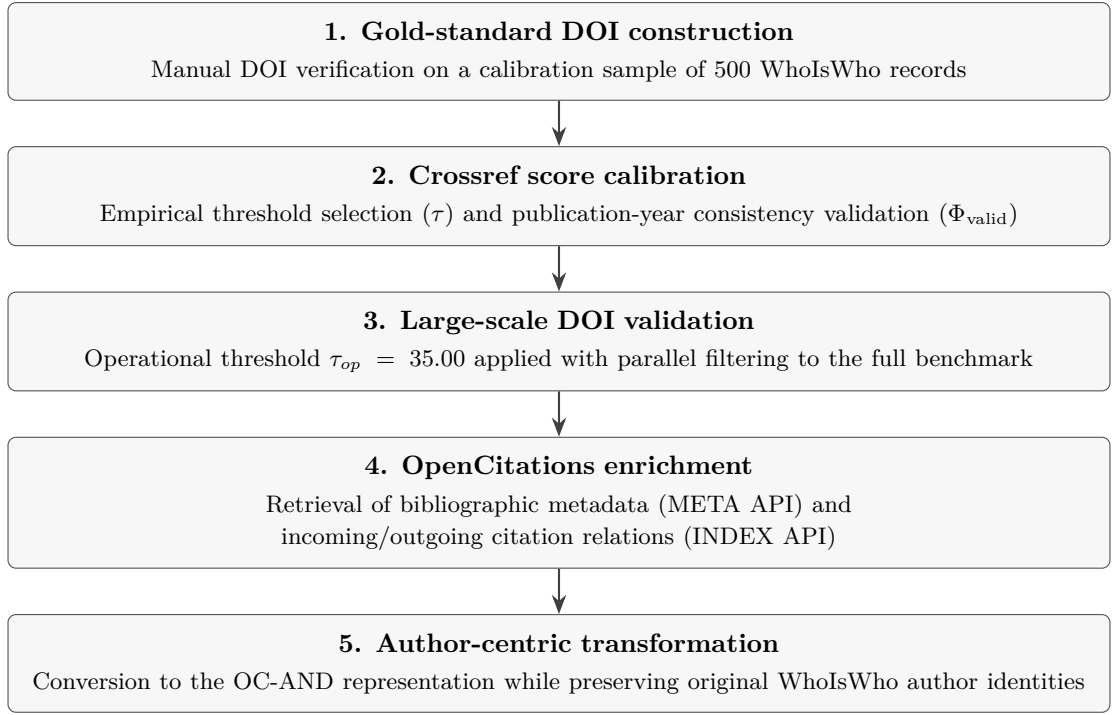
\begin{figure*}[t]
\centering
\begin{tikzpicture}[
    node distance=0.6cm,
    stage/.style={
        rectangle, 
        draw=black!70, 
        fill=black!3, 
        rounded corners=3pt, 
        align=center,
        text width=0.88\textwidth, 
        inner sep=7pt,
        font=\small
    },
    arrow/.style={-{Stealth[scale=1.1]}, thick, draw=black!75}
]
\node (s1) [stage] {\textbf{1. Gold-standard DOI construction}\\[2pt]
\footnotesize Manual DOI verification on a calibration sample of 500 WhoIsWho records};

\node (s2) [stage, below=of s1] {\textbf{2. Crossref score calibration}\\[2pt]
\footnotesize Empirical threshold selection ($\tau$) and publication-year consistency validation ($\Phi_{\text{valid}}$)};

\node (s3) [stage, below=of s2] {\textbf{3. Large-scale DOI validation}\\[2pt]
\footnotesize Operational threshold $\tau_{op}=35.00$ applied with parallel filtering to the full benchmark};

\node (s4) [stage, below=of s3] {\textbf{4. OpenCitations enrichment}\\[2pt]
\footnotesize Retrieval of bibliographic metadata (META API) and incoming/outgoing citation relations (INDEX API)};

\node (s5) [stage, below=of s4] {\textbf{5. Author-centric transformation}\\[2pt]
\footnotesize Conversion to the OC-AND representation while preserving original WhoIsWho author identities};

\draw [arrow] (s1) -- (s2);
\draw [arrow] (s2) -- (s3);
\draw [arrow] (s3) -- (s4);
\draw [arrow] (s4) -- (s5);
\end{tikzpicture}
\caption{Overview of the Bridge4AND dataset construction pipeline. The workflow consists of five stages: (1) gold-standard preparation and DOI verification; (2) Crossref score calibration and publication-year consistency validation; (3) large-scale DOI validation; (4) bibliographic and citation enrichment through OpenCitations; and (5) conversion to the author-centric format.}
\label{fig:pipeline}
\end{figure*}

The construction of the OC-AND dataset from the legacy WhoIsWho benchmark~\cite{chen2023whoiswho} is formalised as Bridge4AND, a five-stage curation and enrichment workflow illustrated in Figure~\ref{fig:pipeline}. A detailed operational representation of each stage, including input/output data artifacts and processing scripts, is provided in Figure~\ref{fig:Bridge4AND_pipeline_detailed}. The objective of the pipeline is to augment the benchmark with open metadata while preserving reproducibility and structural consistency.

\definecolor{s1col}{RGB}{28, 78, 128}   
\definecolor{s2col}{RGB}{0, 115, 115}   
\definecolor{s3col}{RGB}{45, 125, 65}   
\definecolor{s4col}{RGB}{175, 95, 20}   
\definecolor{s5col}{RGB}{105, 45, 135}  

\definecolor{filebg}{RGB}{248, 249, 251}
\definecolor{fileborder}{RGB}{185, 192, 202}

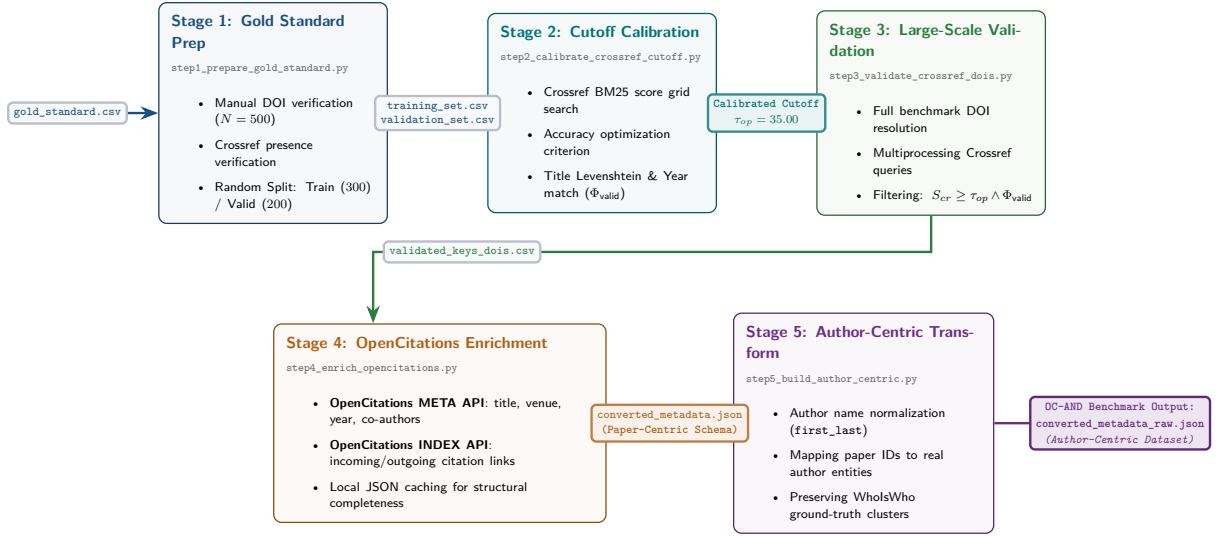
\begin{figure*}[t]
\centering
\resizebox{\textwidth}{!}{%
\begin{tikzpicture}[
    font=\sffamily,
    >=Stealth,
    node distance=1.6cm and 2.2cm,
    stagebox/.style={
        rectangle,
        rounded corners=5pt,
        draw=#1!80!black,
        fill=#1!4,
        thick,
        inner sep=8pt,
        text width=4.5cm,
        align=left
    },
    flowlabel/.style={
        rectangle,
        rounded corners=3pt,
        draw=fileborder,
        fill=filebg,
        font=\fontsize{7.5}{9}\selectfont\ttfamily,
        align=center,
        inner sep=3.5pt
    },
    flowarrow/.style={
        ->,
        >={Stealth[scale=1.15]},
        line width=1.5pt,
        color=#1
    }
]


\node (s1) [stagebox=s1col] {
    {\small\bfseries\textcolor{s1col}{\textbf{Stage 1: Gold Standard Prep}}}\\[1pt]
    {\fontsize{7}{8.5}\selectfont\ttfamily\textcolor{black!60}{step1\_prepare\_gold\_standard.py}}\\[6pt]
    {\fontsize{8}{10.5}\selectfont
    \begin{itemize}\setlength{\itemsep}{1pt}\setlength{\leftmargin}{10pt}
        \item Manual DOI verification ($N=500$)
        \item Crossref presence verification
        \item Random Split: Train ($300$) / Valid ($200$)
    \end{itemize}}
};

\node (s2) [right=of s1, stagebox=s2col] {
    {\small\bfseries\textcolor{s2col}{\textbf{Stage 2: Cutoff Calibration}}}\\[1pt]
    {\fontsize{7}{8.5}\selectfont\ttfamily\textcolor{black!60}{step2\_calibrate\_crossref\_cutoff.py}}\\[6pt]
    {\fontsize{8}{10.5}\selectfont
    \begin{itemize}\setlength{\itemsep}{1pt}\setlength{\leftmargin}{10pt}
        \item Crossref BM25 score grid search
        \item Accuracy optimization criterion
        \item Title Levenshtein \& Year match ($\Phi_{\text{valid}}$)    \end{itemize}}
};

\node (s3) [right=of s2, stagebox=s3col] {
    {\small\bfseries\textcolor{s3col}{\textbf{Stage 3: Large-Scale Validation}}}\\[1pt]
    {\fontsize{7}{8.5}\selectfont\ttfamily\textcolor{black!60}{step3\_validate\_crossref\_dois.py}}\\[6pt]
    {\fontsize{8}{10.5}\selectfont
    \begin{itemize}\setlength{\itemsep}{1pt}\setlength{\leftmargin}{10pt}
        \item Full benchmark DOI resolution
        \item Multiprocessing Crossref queries
        \item Filtering: $S_{cr} \ge \tau_{op} \land \Phi_{\text{valid}}$
    \end{itemize}}
};


\node (s4) [below=2.2cm of s1, anchor=north west, stagebox=s4col, text width=6.8cm] {
    {\small\bfseries\textcolor{s4col}{\textbf{Stage 4: OpenCitations Enrichment}}}\\[1pt]
    {\fontsize{7}{8.5}\selectfont\ttfamily\textcolor{black!60}{step4\_enrich\_opencitations.py}}\\[6pt]
    {\fontsize{8}{10.5}\selectfont
    \begin{itemize}\setlength{\itemsep}{1pt}\setlength{\leftmargin}{10pt}
        \item \textbf{OpenCitations META API}: title, venue, year, co-authors
        \item \textbf{OpenCitations INDEX API}: incoming/outgoing citation links
        \item Local JSON caching for structural completeness
    \end{itemize}}
};

\node (s5) [right=2.8cm of s4, stagebox=s5col, text width=5.2cm] {
    {\small\bfseries\textcolor{s5col}{\textbf{Stage 5: Author-Centric Transform}}}\\[1pt]
    {\fontsize{7}{8.5}\selectfont\ttfamily\textcolor{black!60}{step5\_build\_author\_centric.py}}\\[6pt]
    {\fontsize{8}{10.5}\selectfont
    \begin{itemize}\setlength{\itemsep}{1pt}\setlength{\leftmargin}{10pt}
        \item Author name normalization (\texttt{first\_last})
        \item Mapping paper IDs to real author entities
        \item Preserving WhoIsWho ground-truth clusters
    \end{itemize}}
};


\draw [flowarrow=s1col] ($(s1.west)+(-1.2,0)$) -- (s1.west) 
    node[midway, flowlabel, anchor=east, xshift=-2pt] {gold\_standard.csv};

\draw [flowarrow=s1col] (s1.east) -- (s2.west) 
    node[midway, flowlabel] {training\_set.csv\\validation\_set.csv};

\draw [flowarrow=s2col] (s2.east) -- (s3.west) 
    node[midway, flowlabel, draw=s2col!80, fill=s2col!8] {\textbf{Calibrated Cutoff}\\[1pt]\textcolor{s2col!90!black}{$\tau_{op} = 35.00$}};

\draw [flowarrow=s3col] (s3.south) -- ++(0,-0.8) -| ($(s4.north west)!0.3!(s4.north east)$)
    node[pos=0.5, flowlabel, anchor=west, xshift=6pt] {validated\_keys\_dois.csv};

\draw [flowarrow=s4col] (s4.east) -- (s5.west) 
    node[midway, flowlabel, draw=s4col!80, fill=s4col!8] {converted\_metadata.json\\[1pt]\textbf{(Paper-Centric Schema)}};

\draw [flowarrow=s5col] (s5.east) -- ++(1.4,0) 
    node[midway, flowlabel, anchor=west, xshift=2pt, draw=s5col!90!black, fill=s5col!12] {
        \textbf{OC-AND Benchmark Output:}\\[2pt]
        \textcolor{s5col!90!black}{\textbf{converted\_metadata\_raw.json}}\\[1pt]
        \textit{(Author-Centric Dataset)}
    };

\end{tikzpicture}
}
\caption{Operational workflow of the Bridge4AND pipeline. The top row illustrates DOI retrieval, threshold calibration ($\tau_{op}=35.00$), and validation. The bottom row details metadata and citation enrichment via OpenCitations APIs and final transformation into the author-centric OC-AND benchmark.}
\label{fig:Bridge4AND_pipeline_detailed}
\end{figure*}

\subsection*{Stage 1: Gold standard construction and DOI recovery}
A major structural limitation of the WhoIsWho benchmark is the total absence of Digital Object Identifiers (DOIs). The original schema relies strictly on localized textual strings, making deterministic cross-registry matching impossible without a calibration baseline.

To overcome this, we executed a systematic manual verification workflow on a sample of $N = 500$ publication instances drawn from the high-ambiguity subsets of the WhoIsWho validation partition. For each record, annotators queried the AMiner portal directly to manage structural translations or anglicised normalisations of non-Western titles. Where a DOI was not present in AMiner, secondary queries were executed across open publisher portals. Out of the 500 annotated publications, active DOIs were confirmed for 410 records, while the remaining 90 instances could not be retrieved across persistent registries and were designated \texttt{None}.

The resulting curated set ($\mathrm{DOI}_{\mathrm{gold}}$) was formalized into an auditable tabular schema (\url{gold\_standard.csv}). To ensure data integrity across registries, we queried the Crossref \texttt{/works/\{doi\}} REST endpoint for each entry, appending an explicit status indicator (\texttt{ID\_on\_Crossref}). Finally, the full sample of $N = 500$ records was partitioned using a fixed random seed of 42 into a training split ($N_{\text{train}} = 300$, \texttt{training\_set.csv}) reserved for threshold calibration, and a validation split ($N_{\text{valid}} = 200$, \texttt{validation\_set.csv}).

\subsection*{Stage 2: Crossref score optimisation and preliminary metadata validation}

With the gold standard in place, we established an automated policy for accepting or rejecting DOIs ($DOI_{cr}$) returned by programmatic queries to the Crossref REST API~\cite{hendricks2020crossref}.

Crossref BM25~\cite{elastic_bm25} queries are constructed using the tuple $(T, Y)$, where $T$ denotes 
the normalised publication title and $Y$ denotes the publication year. BM25 is 
a probabilistic ranking function that combines term frequency, inverse document 
frequency, and document length normalization~\cite{robertson2009probabilistic}. 
The Crossref API returns candidate 
matches ranked by relevance score $S_{\text{cr}}$ using BM25. Since $S_{\text{cr}}$ 
is query-dependent, it must be calibrated empirically before being used as a 
classification threshold.
 
Let $\tau \in \mathbb{R}^+$ be a candidate cutoff threshold. We execute a grid search over the 
discrete values of $S_{\text{cr}}$ observed in a training split of the gold standard, selecting 
the optimal threshold $\tau_{\text{opt}}$ to maximise classification accuracy:
 
$$\tau_{\text{opt}} = \arg\max_{\tau} \text{Accuracy}(\tau) = \arg\max_{\tau} \frac{\text{TP}(\tau) + \text{TN}(\tau)}{\text{TP}(\tau) + \text{TN}(\tau) + \text{FP}(\tau) + \text{FN}(\tau)}$$
 
where $\text{TP}(\tau)$ denotes true positives (correct DOI matches), $\text{TN}(\tau)$ denotes 
true negatives (correctly rejected candidates), $\text{FP}(\tau)$ denotes false positives (spurious matches where a high-ranking candidate is actually a different publication), and $\text{FN}(\tau)$ denotes false negatives (valid DOIs rejected by the threshold). Accuracy is adopted as the optimisation criterion because false positives introduce noise into the citation graph, whereas false negatives can be retrieved through retry phases.
 
The score threshold alone is insufficient to guarantee correct DOI assignment. Using Crossref API, structurally 
similar records, such as book reviews, corrigenda, or alternative editions, may obtain high 
relevance scores despite referring to different publications. To reduce such errors, we introduce 
a lightweight metadata validation function $\Phi_{\text{valid}}$ that acts as a secondary 
consistency check.
 
A retrieved DOI is accepted if and only if both conditions hold simultaneously:
 
$$\text{DOI}_{\text{Accepted}} \Leftrightarrow (S_{\text{cr}} \geq \tau_{\text{opt}}) \wedge \Phi_{\text{valid}}(M_{\text{json}}, M_{\text{cr}})$$
 
where $M_{\text{json}}$ represents the metadata object of the source query paper extracted from 
the legacy dataset, and $M_{\text{cr}}$ represents the candidate metadata schema returned by 
the Crossref API response.
 
In the present implementation, $\Phi_{\text{valid}}$ is formulated as a boolean function that 
relies exclusively on publication year agreement:
 
$$\Phi_{\text{valid}}(M_{\text{json}}, M_{\text{cr}}) = \mathbb{I}(Y_{\text{json}} = Y_{\text{cr}})$$
 
where $Y_{\text{json}}$ and $Y_{\text{cr}}$ denote the publication year extracted from 
$M_{\text{json}}$ and $M_{\text{cr}}$ respectively, and $\mathbb{I}(\cdot)$ is the indicator 
function returning 1 if the condition holds and 0 otherwise. Publication year was selected 
because it is the only metadata field available with near-complete coverage and consistent 
reliability across both source and target infrastructures. In principle, richer validation 
schemes could incorporate additional metadata attributes, but introducing partially populated 
or inconsistently curated fields would risk rejecting otherwise correct DOI assignments. The year-based validator therefore provides a conservative and broadly applicable baseline, while leaving room for more expressive metadata-based validation strategies when additional reliable information is available.

\subsection*{Stage 3: Large-scale DOI validation}
The threshold $\tau_{opt}$ and the validator $\Phi_{valid}$ are applied to the full WhoIsWho dataset. Candidate DOI assignments are verified against the Crossref REST API and partitioned into three disjoint outputs corresponding to validated records, rejected matches, and unresolved cases (\texttt{validated\_keys\_dois.csv}, \texttt{rejected\_items.csv}, and \texttt{error\_items.csv}).
Only publications in \texttt{validated\_keys\_dois.csv} proceed to Stage~4.

\subsection*{Stage 4: OpenCitations metadata and citation enrichment}

Verified DOI mappings are enriched via two complementary OpenCitations APIs (accessed as of January 2025)~\cite{peroni2020opencitations}:
\begin{itemize}
  \item The \textbf{OpenCitations Meta API}~\cite{opencitations_meta_api} (\url{https://api.opencitations.net/meta/v1/metadata/}), providing bibliographic metadata: publication titles,publication venue (including raw serial identifiers such as ISSN and OMID), publication year, co-author lists.
  \item The \textbf{OpenCitations Index API}~\cite{opencitations_index_api} (\url{https://api.opencitations.net/index/v2/}), querying two distinct operations to capture the citation neighborhood:
  \begin{itemize}
    \item \texttt{/citations/\{doi\}} for incoming citations;
    \item \texttt{/references/\{doi\}} for outgoing references.
  \end{itemize}
\end{itemize}

Publications whose validated DOI has no record in OpenCitations Meta are excluded from the resulting publication files (\texttt{train\_pub.json} and \texttt{sna\_valid\_pub.json}).
Unpopulated properties, such as missing abstracts, non-existing affiliation, or vacant citation nodes—are systematically retained as empty string literals (\texttt{""}) or empty arrays (\texttt{[]}) to ensure structural consistency. Conversely, since OpenCitations Meta does not natively provide topical keywords, we synthetically derive keywords directly from the publication title: after removing common English stopwords, salient alphabetic terms longer than three characters are extracted (up to a maximum of 15 terms per record) to preserve schema compatibility with downstream disambiguation architectures.

\subsection*{Stage 5: Author-centric transformation}
The final stage converts the enriched paper-centric schema into the author-centric representations required by graph-based AND frameworks.

The supervised label files (\texttt{train\_author.json} and \texttt{sna\_valid\_ground\_truth.json}) are distributed unchanged from WhoIsWho na-v3, preserving original cluster memberships and the training/validation split structure. They therefore also reference WhoIsWho publications for which no OpenCitations record was retrieved; the effective ground truth for OC-AND is the restriction of these clusters to the publications included in the corresponding publication file. No inference input is provided for the training partition: following the WhoIsWho convention, training publications are paired with target names through \texttt{train\_author.json}, i.e.\ by identifier. The unlabelled validation input (\texttt{sna\_valid\_raw.json}) is instead constructed by inverting the paper-centric schema into a mapping from every normalised author name string occurring in OpenCitations Meta co-author lists to flat lists of publication identifiers.

Author strings are processed through a deterministic normalization sequence (lowercase conversion, abbreviation expansion, punctuation stripping, and extraction of first and last name tokens only) to align with the conventions assumed by the WhoIsWho disambiguation pipeline~\cite{chen2023whoiswho}.

\section*{Data Records}
\label{sec:data_records}
The OC-AND dataset is distributed as five separate JSON files organized into a training partition and a validation partition, summarized in Table~\ref{tab:files}.

\begin{table}[htbp]
\centering
\caption{Summary and partition mapping of the distributed OC-AND dataset files.}
\label{tab:files}
\resizebox{0.95\textwidth}{!}{%
\begin{tabular}{llcl}
\toprule
\textbf{Partition} & \textbf{File Name} & \textbf{Format} & \textbf{Description} \\
\midrule
Training   & \texttt{train\_pub.json}       & JSON & Enriched bibliographic records (paper-centric) \\
    & \texttt{train\_author.json}    & JSON & Ground-truth clusters (WhoIsWho na-v3, unchanged) \\
\addlinespace
Validation & \texttt{sna\_valid\_pub.json}  & JSON & Enriched bibliographic records (paper-centric) \\
    & \texttt{sna\_valid\_raw.json}  & JSON & Unlabelled inference input, keyed by OpenCitations author strings \\
    & \texttt{sna\_valid\_ground\_truth.json} & JSON & Ground-truth clusters (WhoIsWho na-v3, unchanged) \\
\bottomrule
\end{tabular}%
}
\end{table}
 
\subsection*{Publication Records}

\texttt{train\_pub.json} and \texttt{sna\_valid\_pub.json} share an identical paper-centric schema enumerated in Table~\ref{tab:pub_schema}. Fields not present in OpenCitations (abstract, author affiliations) are created as empty strings or empty arrays to maintain schema compatibility, whereas keywords are extracted from publication titles by pipeline design. An example publication record is shown in Figure~\ref{fig:pub_example}.

\begin{table}[htbp]
\centering
\caption{Field-level schema for paper-centric publication records.}
\label{tab:pub_schema}
\small
\begin{tabular}{llp{8.5cm}}
\toprule
\textbf{Field} & \textbf{Type} & \textbf{Description} \\
\midrule
\texttt{id} & string & Internal publication identifier. \\
\texttt{title} & string & Publication title. \\
\texttt{abstract} & string & Abstract text, or empty string. \\
\texttt{keywords} & array & Topical descriptors, or empty array. \\
\texttt{authors} & array & Co-author objects (\texttt{name}, \texttt{org}). \\
\texttt{venue} & string & Publication venue (text string). \\
\texttt{year} & integer & Publication year . \\
\texttt{doi} & string & DOI identifier (verified). \\
\texttt{outgoing\_citations} & array & Available Identifiers of cited works [OMID, DOI, OpenAlex ID,Pmid]. \\
\texttt{incoming\_citations} & array & Available Identifiers of citing works [OMID, DOI, OpenAlex ID,Pmid]. \\
\texttt{outgoing\_citations\_count} & integer & Count of outgoing links. \\
\texttt{incoming\_citations\_count} & integer & Count of incoming links. \\
\bottomrule
\end{tabular}
\end{table}

\begin{figure}[!h]
\centering
\footnotesize
\begin{verbatim}
{
  "AjhViKPu": {
    "id": "AjhViKPu",
    "title": "Linking stoichiometric homeostasis with 
              ecosystem structure, functioning, and stability",
    "abstract": "",
    "keywords": [
      "linking", "stoichiometric", "homeostasis", 
      "ecosystem", "structure", "functioning", "stability"
    ],
    "authors": [
      {"name": "qiang yu", "org": ""},
      {"name": "quansheng chen", "org": ""},
      {"name": "j elser", "org": ""}
    ],
    "venue": "Nature Precedings",
    "year": 2010,
    "doi": "10.1038/npre.2010.5255.1",
    "outgoing_citations": [],
    "incoming_citations": ["omid:br/06130716788 10.1016/..."],
    "outgoing_citations_count": 0,
    "incoming_citations_count": 1
  }
}
\end{verbatim}
\caption{Example publication record from OC-AND. Abstract and author affiliations are empty per OpenCitations availability.}
\label{fig:pub_example}
\end{figure}

\subsection*{Ground-Truth Cluster Files}

\texttt{train\_author.json} and \texttt{sna\_valid\_ground\_truth.json} share an identical author-centric structure: normalized author name strings map to cluster objects where cluster IDs are keys and publication ID lists are values. Both files are identical to their WhoIsWho na-v3 counterparts; publication identifiers without a record in the OC-AND publication files are not removed. An example is provided in Figure~\ref{fig:author_example}.

\begin{figure}[!h]
\centering
\footnotesize
\begin{verbatim}
{
  "haifeng_qian": {
    "GHJ7aB2u": ["SYO964qr", "XfIPP9AU", "pd9Aa57C", 
                 "3DKK79ow", "cKEedrnU"],
    "Pg3Cdsmd": ["aKT0W4tz", "ZO9dg5oq", "m6YVaacR"]
  }
}
\end{verbatim}
\caption{Ground-truth structure. Multiple cluster IDs per author encode genuine ambiguity; publication IDs within each cluster belong to the same real author identity.}
\label{fig:author_example}
\end{figure}

\subsection*{Validation Inference Input}

\texttt{sna\_valid\_raw.json} is the input to be disambiguated. For each author name, it lists the identifiers of all publications in which that name appears in the OpenCitations Meta author list, without indicating which real person each publication belongs to. Unlike the ground truth, it is keyed by all author strings occurring in these publications, co-authors included, of which the 80 target names are a subset. Name variants are not merged (e.g.\ \texttt{haifeng\_qian} and \texttt{hai-feng\_qian} are distinct keys), and publications with an empty author list in OpenCitations Meta are not indexed under any key. As a consequence, the publications listed under a target name do not fully coincide with those assigned to it in the ground truth: some ground-truth publications appear only under a variant string or under no key. Conversely, publications co-authored by two target names are listed under both in the inference input, whereas the ground truth assigns them to a cluster of only one of the two. An example is shown in Figure~\ref{fig:raw_example}.

\begin{figure}[!h]
\centering
\footnotesize
\begin{verbatim}
{
  "haifeng_qian":  ["nVZXR80K", "H10UXbFb", "rlSCezs9",
                    "m6YVaacR", ...],
  "hai-feng_qian": ["2vbP7GtV", "lL8jsJgX", ...],
  "xiaonan_wang":  ["MDFjSzCn"]
}
\end{verbatim}
\caption{Example of the validation inference input. Keys are author strings from OpenCitations Meta, including co-authors and name variants of target names; cluster labels are withheld.}
\label{fig:raw_example}
\end{figure}

\section*{Dataset Profile and Properties}
\label{sec:dataset_profile}
The execution of the Bridge4AND pipeline transforms the informational layout of the WhoIsWho benchmark, resulting in a metadata sparsity characteristic of open scholarly infrastructures.

\subsection*{Metadata Completeness and Richness Profile}
The structural divergence between closed data silos and open bibliographic paradigms is quantified via a field-level completeness audit in Table~\ref{tab:metadata_completeness}. 
 For the venue field, OpenCitations Meta returns composite text strings that frequently embed persistent publication identifiers (such as ISSN and OMID). To preserve the raw state of the metadata, our pipeline retains these composite venue strings verbatim without executing secondary normalization or entity reconciliation across records.

\begin{table}[htbp]
\centering
\caption{Metadata field availability profile: field-level comparison.}
\label{tab:metadata_completeness}
\small
\begin{tabular}{lcc}
\toprule
\textbf{Metadata Field} & \textbf{WhoIsWho} & \textbf{OC-AND} \\
\midrule
Title & \checkmark & \checkmark \\
Authors (Co-author list) & \checkmark & \checkmark \\
Publication Year & \checkmark & \checkmark \\
Venue & \checkmark & \checkmark \\
Keywords  & \checkmark & \checkmark\\
\midrule
Abstract & \checkmark & $\times$ \\
Author Affiliations (\textit{org}) & \checkmark & $\times$ \\
\midrule
Digital Object Identifier (DOI) & $\times$ & \checkmark \\
Incoming Citations & $\times$ & \checkmark \\
Outgoing Citations & $\times$ & \checkmark \\
\bottomrule
\end{tabular}
\end{table}

The transition demonstrates a clear trade-off. The legacy WhoIsWho dataset relies on dense, proprietary metadata text, exposing high coverage for abstracts (approximately 78.27\%) and institutional affiliations (approximately 72.79\%). In OC-AND, these fields are absent, reflecting structural limitations across open registries in terms of what metadata is routinely captured and indexed. This textual sparsity is counterbalanced by a substantial relational layer: citation links are populated for over 92.9\% of all records.

\subsection*{Volumetric and Partition Structure}

The ground-truth files of OC-AND are identical to those of WhoIsWho na-v3 and all target author names are retained in both partitions. The two datasets differ in their publication files and in the validation inference input, as summarised in Table~\ref{tab:dataset_comparison_full}. The first block traces the reduction of publications through the pipeline: publications enter the OC-AND publication files only if a DOI was validated against Crossref (Stage~3) and an OpenCitations record was retrieved for it (Stage~4). Publication instances count author--paper assignments and exceed publications because some papers include two target names among their co-authors. The second block describes the ground-truth clusters that remain when restricted to the publications in each publication file; clusters with no remaining publication disappear. The third block concerns the validation inference input only, since no inference input exists for the training partition. Its keys include all author strings found in OpenCitations Meta, co-authors and name variants included, and the last row counts the publications listed under the 80 target names (see Validation Inference Input).

\begin{table}[htbp]
\centering
\caption{Differences between WhoIsWho na-v3 and OC-AND by partition. Cluster statistics refer to the ground truth restricted to the publications in each publication file.}
\label{tab:dataset_comparison_full}
\small
\begin{tabular}{lccccc}
\toprule
 & \multicolumn{2}{c}{\textbf{Training}} & & \multicolumn{2}{c}{\textbf{Validation}} \\
\cline{2-3} \cline{5-6}
\rule{0pt}{3ex}\textbf{Metric} & \textbf{WhoIsWho} & \textbf{OC-AND} & & \textbf{WhoIsWho} & \textbf{OC-AND} \\
\midrule
\multicolumn{6}{l}{\textit{Publications}} \\
With validated DOI (Stage 3) & -- & 46,738 & & -- & 28,930 \\
In publication file (Stage 4) & 73,835 & 43,640 & & 46,287 & 27,265 \\
Publication instances & 74,117 & 43,858 & & 46,367 & 27,321 \\
\midrule
\multicolumn{6}{l}{\textit{Ground-truth clusters}} \\
Clusters & 2,595 & 2,152 & & 1,612 & 1,324 \\
Mean publications per cluster & 28.56 & 20.38 & & 28.76 & 20.64 \\
Median publications per cluster & 12 & 8 & & 12 & 7 \\
Maximum publications per cluster & 602 & 485 & & 456 & 384 \\
95th percentile cluster size & 115.30 & 89.00 & & 112.00 & 87.85 \\
Mean real authors per name & 16.22 & 13.45 & & 20.15 & 16.55 \\
Maximum real authors per name & 96 & 77 & & 74 & 63 \\
\midrule
\multicolumn{6}{l}{\textit{Inference input}} \\
Author name keys & -- & -- & & 80 & 53,211 \\
Publications under target names & -- & -- & & 46,287 & 22,338 \\
\bottomrule
\end{tabular}
\end{table}
\section*{Technical Validation}
\label{sec:technical_validation}
Technical validation addresses the accuracy of the automated DOI assignment pipeline and the suitability of the dataset for AND evaluation.

\subsection*{DOI assignment accuracy}
An audit of the $N=500$ gold standard records reveals distinct constraints on open metadata transfer: 82.0\% (410 records) were manually successfully resolved to active DOIs, while 18.0\% (90 records) proved unresolvable. Linguistic boundaries emerged as the primary factor in unavailability: 98 records carried titles exclusively in Chinese, of which 76 possessed valid DOIs entirely absent from Crossref indexes.

\begin{figure}[ht]
\centering
\includegraphics[width=0.9\columnwidth]{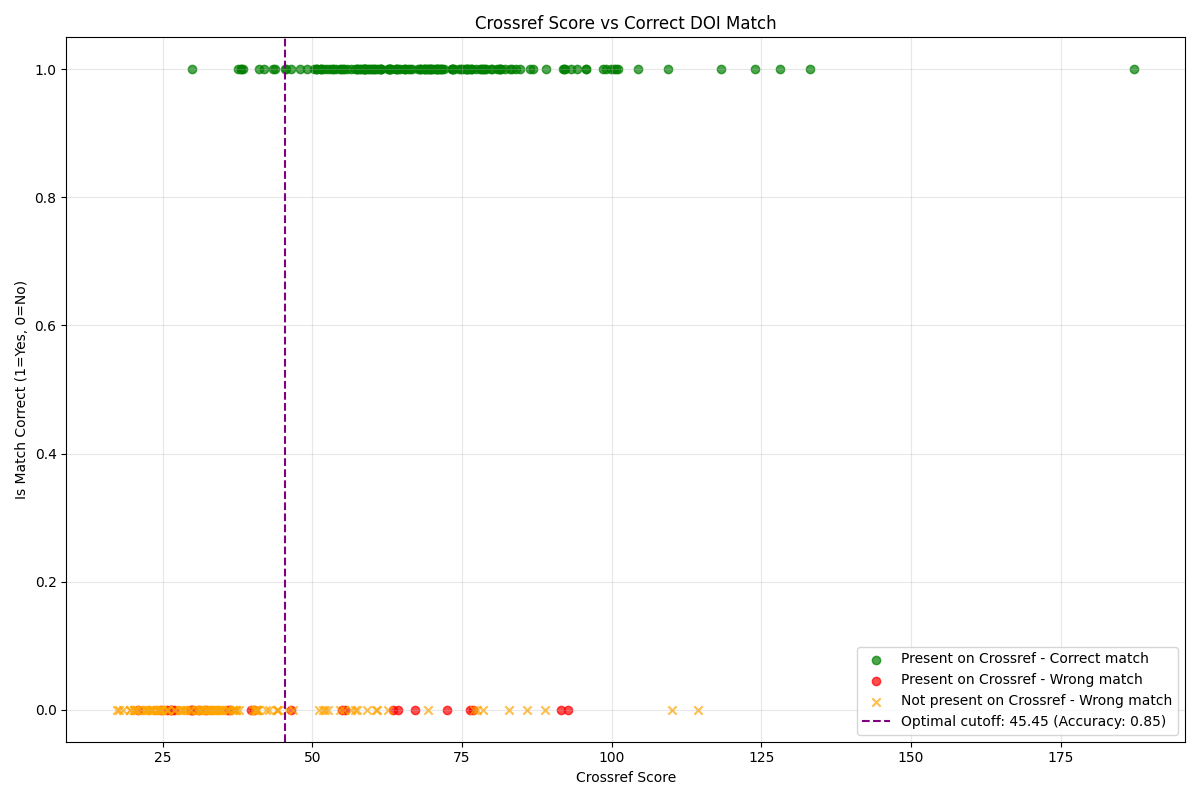}
\caption{Empirical distributions of Crossref relevance scores ($S_\mathrm{cr}$) on the calibration split, mapped against binary ground truth status. The dashed vertical line isolates the computed optimal $\tau_\mathrm{opt}$ boundary, whereas the subsequent operational pipeline threshold is set at $\tau = 35$.}
\label{fig:crossref_score_analysis}
\end{figure}

Table~\ref{tab:cutoff_comparison} reports the performance of the empirical thresholds on the calibration split, whose score distributions are illustrated in Figure~\ref{fig:crossref_score_analysis}, and of the operational configuration on the held-out validation split. On the calibration split, the accuracy-optimal ($\tau_{opt}=45.45$) and $F_1$-optimal ($\tau=41.01$) thresholds yield similar performance.

\begin{table}[!h]
\centering
\small
\caption{Performance of Crossref DOI assignment. Calibration split ($N_{\text{train}}=300$): score threshold only. Held-out validation split ($N_{\text{valid}}=200$): full operational configuration ($\tau_{op}$ and $\Phi_{\text{valid}}$). A positive case is a publication for which Crossref returns the correct DOI.}
\label{tab:cutoff_comparison}
\resizebox{0.95\textwidth}{!}{%
\begin{tabular}{lccccccccc}
\toprule
Criterion & $\tau$ & Acc & F$_1$ & Prec & Recall & TP & FP & TN & FN \\
\midrule
\multicolumn{10}{l}{\textit{Calibration split, score threshold only}} \\
Accuracy-opt & 45.45 & 0.853 & 0.870 & 0.808 & 0.942 & 147 & 35 & 109 & 9 \\
F$_1$-opt & 41.01 & 0.850 & 0.870 & 0.791 & 0.968 & 151 & 40 & 104 & 5 \\
Operational & 35.00 & 0.793 & 0.833 & 0.718 & 0.994 & 155 & 61 & 83 & 1 \\
\midrule
\multicolumn{10}{l}{\textit{Held-out validation split, $\tau_{op}$ and $\Phi_{\text{valid}}$}} \\
\textbf{Operational} & \textbf{35.00} & \textbf{0.865} & \textbf{0.893} & \textbf{0.837} & \textbf{0.958} & \textbf{113} & \textbf{22} & \textbf{60} & \textbf{5} \\
\bottomrule
\end{tabular}%
}
\end{table}

For deployment on the full dataset, we instead adopted a lower operational cutoff of $\tau_{op}=35.00$. On the calibration split, this threshold deliberately favours recall over precision (recall 0.994, precision 0.718), ensuring that valid DOIs with lower relevance scores are not prematurely discarded and leaving the removal of false positives to the subsequent metadata validation layer ($\Phi_{\text{valid}}$). On the held-out validation split, the complete configuration reaches a precision of 0.837 and a recall of 0.958.

\begin{figure}[!h]
\centering
\begin{tikzpicture}[
    font=\small, >=Stealth,
    box/.style={rectangle, draw=black!75, fill=gray!5, rounded corners=3pt, align=center, inner sep=4pt, text width=4.6cm},
    gate/.style={rectangle, draw=black!90, fill=gray!15, rounded corners=3pt, align=center, inner sep=4pt, text width=4.6cm},
    rej/.style={rectangle, draw=black!60, fill=white, rounded corners=2pt, align=center, inner sep=3pt, font=\scriptsize, text width=2.6cm},
    arrow/.style={-{Stealth[scale=0.8]}, semithick, draw=black!80}
]
\node (in) [box] {\textbf{Held-out validation split} ($N=200$)\\[1pt]\scriptsize 118 correct / 82 incorrect candidates};
\node (g1) [gate, below=0.6cm of in] {\textbf{Score threshold}\\[1pt]\scriptsize $S_{cr} \geq \tau_{op} = 35.00$};
\node (r1) [rej, right=0.6cm of g1] {\textbf{Rejected: 46}\\TN 44 \quad FN 2};
\node (g2) [gate, below=0.6cm of g1] {\textbf{Metadata validation}\\[1pt]\scriptsize $\Phi_{\text{valid}}$};
\node (r2) [rej, right=0.6cm of g2] {\textbf{Rejected: 19}\\TN 16 \quad FN 3};
\node (acc) [box, below=0.6cm of g2] {\textbf{Accepted: 135}\\[1pt]\scriptsize TP 113 \quad FP 22};
\draw [arrow] (in) -- (g1);
\draw [arrow] (g1) -- (r1);
\draw [arrow] (g1) -- (g2);
\draw [arrow] (g2) -- (r2);
\draw [arrow] (g2) -- (acc);
\end{tikzpicture}
\caption{Sequential filtering and classification outcomes of the operational configuration on the held-out validation split ($N_{\text{valid}}=200$).}
\label{fig:pipeline_flowchart}
\end{figure}
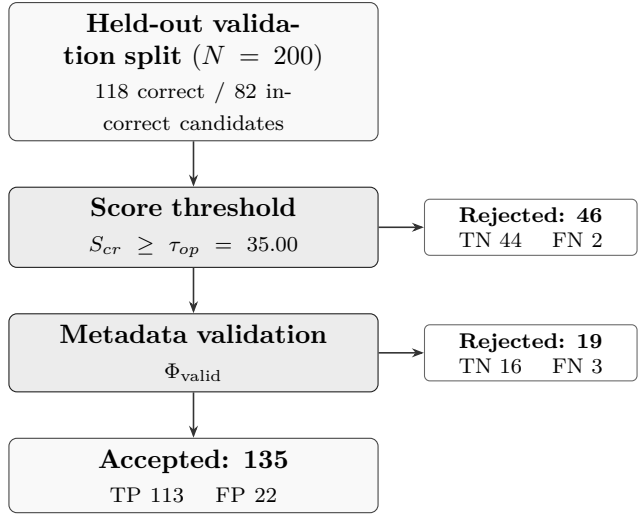

The sequential filtering of the held-out split is mapped in Figure~\ref{fig:pipeline_flowchart}. The score threshold removes 46 candidates (44 incorrect, 2 correct), $\Phi_{\text{valid}}$ removes a further 19 (16 incorrect, 3 correct), and 135 DOIs are accepted, of which 113 are correct. Of the 22 false positives, 13 concern publications for which no DOI exists.

\subsubsection*{Functional validation via disambiguation baseline}
 
We conducted two distinct experiments using the BOND model~\cite{zhang2022bond}, an end-to-end AND system that combines local metric learning via graph auto-encoder with global DBSCAN clustering on multi-relational graphs, to validate the dataset's utility for AND evaluation. Citation topology was disabled in both configurations to isolate the effect of metadata availability.
 
\begin{itemize}
    \item \textbf{Experiment 1 (BOND on WhoIsWho):} Run on the original WhoIsWho partitions to establish our baseline, successfully reproducing published results. Particular emphasis is placed on dissecting the model's behavior regarding splitting errors (where a single author entity is fragmented into multiple clusters) and lumping errors (where distinct authors with similar names are incorrectly merged). Following the evaluation framework proposed in~\cite{cappelli2025eval}, which provides cluster-level reformulations of these error types, this granular evaluation allows us to map the baseline structural weaknesses before moving to open environments.
    \item \textbf{Experiment 2 (BOND on OC-AND):} Run as a drop-in replacement on the OC-AND dataset, verifying that the dataset is fully compatible with existing pipelines without expensive engineering changes.
\end{itemize}
 
Full performance metrics are reported in Table~\ref{tab:results}.

\begin{table}[ht]
\centering
\caption{Performance of the BOND model on the WhoIsWho na-v3 baseline and the OC-AND dataset (text-only configuration, citation topology disabled).}
\label{tab:results}
\small
\begin{tabular}{L{4.2cm}L{3.2cm}cccc}
\toprule
& & \multicolumn{2}{c}{\textbf{WhoIsWho na-v3}} & \multicolumn{2}{c}{\textbf{OC-AND dataset}} \\
\cmidrule(lr){3-4}\cmidrule(lr){5-6}
\textbf{Dimension} & \textbf{Metric} & Train & Valid & Train & Valid \\
\midrule
Pairwise consistency
  & Precision (PWP)   & 0.7149 & 0.8170 & 0.7058 & 0.8003 \\
  & Recall (PWR)      & 0.9546 & 0.9442 & 0.3828 & 0.3650 \\
  & F1 (F1$_\mathrm{PW}$) & 0.8175 & 0.8760 & 0.4964 & 0.5013 \\
\midrule
Instance-level (B$^3$)
  & Precision (B$^3_P$) & 0.8355 & 0.8587 & 0.8612 & 0.8646 \\
  & Recall (B$^3_R$)    & 0.9433 & 0.9404 & 0.3670 & 0.3557 \\
  & F1 (F1$_\mathrm{B^3}$) & 0.8861 & 0.8977 & 0.5147 & 0.5041 \\
\midrule
Structural (cluster-level)
  & Lumping Error (LE)    & 0.1560 & 0.1275 & 0.0175 & 0.0232 \\
  & Splitting Error (SE)  & 0.1333 & 0.1348 & 0.7797 & 0.7865 \\
  & Structural Score      & 0.8553 & 0.8688 & 0.6014 & 0.5951 \\
\midrule
\textbf{Ranking criterion}
  & \textbf{Composite Score} & \textbf{0.8594} & \textbf{0.8854} & \textbf{0.5265} & \textbf{0.5215} \\
\bottomrule
\end{tabular}
\end{table}

\section*{Usage Notes}
\label{sec:usage_notes}

To facilitate the effective reuse, reproduction, and extension of this benchmark, researchers should consider the following operational guidelines:

\subsubsection*{Datasets Split Design}
Because BOND operates as a bootstrap-from-scratch framework, it does not optimize parametric weights or learn global latent representations during a supervised training phase that are subsequently transferred to the validation step. Consequently, the division between the training and validation splits does not serve to mitigate traditional machine learning overfitting, nor does it pose risks of parametric data leakage. Instead, the two partitions function as independent, structurally isolated validation gates to verify the stability and generalizability of disambiguation heuristics across distinct data samples.

The two partitions differ in how publications are paired with target names. In training, pairing is by identifier through the ground-truth file; in validation, it relies on author strings as recorded in OpenCitations Meta, so the inference input additionally reflects name variants and missing author entries (see Validation Inference Input). Validation results therefore include this retrieval effect, which is absent from training.

\subsubsection*{Evaluation Framework and Metrics of the Functional validation}

Clustering quality is assessed using pairwise metrics, instance-level B3 precision and recall, and cluster-level structural error decomposition~\cite{cappelli2025eval}. For each target name, the ground truth is restricted to the publications for which a prediction is produced: publications absent from the publication file, or not listed under the target name in the validation inference input, are excluded from evaluation rather than counted as errors. All metrics are computed per name; pairwise and B3 counts are summed across names, while LE and SE are averaged across names.

\textbf{Pairwise Metrics.} Let $TP$ be the number of publication pairs correctly co-clustered, $FP$ the number incorrectly placed in the same predicted cluster, and $FN$ the number belonging to the same ground-truth cluster but placed in different predicted clusters:
\begin{align}
\text{Pairwise Precision} &= \frac{TP}{TP + FP}\\
\text{Pairwise Recall} &= \frac{TP}{TP + FN}\\
\text{Pairwise F1} &= \frac{2 \cdot \text{Prec} \cdot \text{Rec}}{\text{Prec} + \text{Rec}}
\end{align}
Pairwise F1 evaluates global co-clustering consistency across all publication pairs.

\textbf{B3 Precision and Recall.} For each publication $i$, let $P(i)$ denote its predicted cluster and $T(i)$ its ground-truth cluster:
\begin{align}
\text{B3 Precision} &= \frac{1}{N} \sum_{i=1}^{N} \frac{|P(i) \cap T(i)|}{|P(i)|}\\
\text{B3 Recall} &= \frac{1}{N} \sum_{i=1}^{N} \frac{|P(i) \cap T(i)|}{|T(i)|}
\end{align}
B3 metrics measure cluster correctness at the instance level and are robust to cluster size imbalance, making them particularly suitable for AND where author publication counts vary substantially.

Standard aggregate metrics conflate two qualitatively distinct failure modes. We therefore decompose errors explicitly at the cluster level~\cite{cappelli2025eval}:

For a predicted cluster $C$, let $g(C)$ denote the number of distinct ground-truth authors it contains. For a ground-truth author cluster $T$, let $p(T)$ denote the number of predicted clusters intersecting it.

\textbf{Normalized Lumping Error (LE):}
\begin{align}
\text{LE} = \frac{1}{|\mathcal{C}_P|} \sum_{C \in \mathcal{C}_P} \frac{\max(0, g(C) - 1)}{g(C)}
\end{align}
LE quantifies cluster contamination: it measures what fraction of predicted clusters incorrectly merge multiple ground-truth authors. An LE of 0 indicates all predicted clusters are pure; as LE approaches 1, predicted clusters increasingly mix distinct authors.

\textbf{Normalized Splitting Error (SE):}
\begin{align}
\text{SE} = \frac{1}{|\mathcal{C}_T|} \sum_{T \in \mathcal{C}_T} \frac{\max(0, p(T) - 1)}{p(T)}
\end{align}
SE quantifies author fragmentation: it measures what fraction of ground-truth authors are scattered across multiple predicted clusters. An SE of 0 indicates all ground-truth authors are recovered completely; as SE approaches 1, authors are increasingly over-partitioned.

These structural metrics are conceptually independent from B3: they operate on cluster topology rather than instance-level overlap, revealing whether disambiguation failures stem from aggressive merging (high LE) or conservative fragmentation (high SE). 

For open scholarly infrastructures, lumping errors carry higher operational cost: they introduce silent contamination in author profiles that is difficult to detect and correct downstream, whereas splitting errors are often recoverable through complementary signals.

\textbf{Structural Score and Composite Ranking.} The two error components are aggregated into:
\begin{align}
S_{\text{struct}} = 1 - \frac{\text{LE} + \text{SE}}{2}
\end{align}
A \textbf{Composite Score} then integrates all signals:
\begin{align}
\text{Composite Score} = 0.3 \cdot \text{F1}_{\text{PW}} + 0.5 \cdot \text{F1}_{\text{B3}} + 0.2 \cdot S_{\text{struct}}
\end{align}
The Composite Score is used as a compact ranking criterion for model selection. Full methodological details, including sensitivity analysis of weighting schemes and validation on synthetic case studies, are presented in~\cite{cappelli2025eval}.

\subsubsection*{Metadata Sparsity and Model Performance}As quantified in Table~\ref{tab:results}, transitioning to open metadata ecosystems is associated with pronounced metadata-level sparsity. The observed increase in splitting error (with the metric rising from $\text{SE} = 0.1348$ to $0.7865$ on the validation split) may be explained by the reduced availability of rich, co-occurring textual features. This behavioral shift suggests that downstream clustering models tend toward fragmentation when deprived of dense metadata contexts, warranting further investigation into whether alternative signals can compensate for the lost feature richness and preserve clustering cohesion.

Since the ground truth in the functional validation is restricted to the publications for which a prediction is produced, the lower recall does not reflect publications missing from OC-AND. The splitting error is also comparable on the training partition ($\text{SE} = 0.7797$), where publications are paired with target names by identifier, which suggests that the increase is not primarily driven by the name-string retrieval of the validation input.

\subsubsection*{Exploiting Citation Topology}
The \texttt{outgoing\_citations} and \texttt{incoming\_citations} fields encode the citation network directly into the publication records. This allows researchers to explore whether graph topology and bibliographic coupling can compensate or mitigate the loss of missing text-rich features.

\section*{Conclusions}
\label{sec:conclusions}
We introduced OC-AND, an open citation-enriched benchmark for AND constructed by re-mapping the WhoIsWho benchmark onto bibliographic and citation information retrieved from OpenCitations. The Bridge4AND pipeline provides a reproducible procedure for DOI recovery, Crossref-based validation, OpenCitations enrichment, and author-centric transformation while preserving the original ground-truth files and partition structure, and restricting publications records to those resolvable through Crossref and OpenCitations. The resulting dataset substantially changes the information available to AND systems: abstracts, and affiliations are absent from the reconstructed representation, while DOI identifiers are sistematically added and citation relations are introduced for the large majority of records. The BOND experiments show that this change produces a substantially different and more challenging disambiguation environment while remaining compatible with an existing AND pipeline. Importantly, the present experiments do not evaluate the benefit of using citation topology for disambiguation, since citation information was disabled in the functional validation. OC-AND is therefore intended as an open evaluation resource on which citation-aware AND methods can be assessed in future work.

%
\section*{Data Availability}
The OC-AND dataset is deposited at OC-AND-benchmark and is available at \url{https://drive.google.com/drive/folders/1kAHMOCEQhx9N9VS2kInyN9A6rY6F2ajU?usp=sharing} .

\section*{Code Availability}
The Bridge4AND pipeline scripts are available at \url{https://github.com/fra533/Bridge4AND.git}. The pipeline is implemented in Python 3 (\texttt{requests}, \texttt{pandas}, \texttt{python-Levenshtein}).
Cite: Cappelli, F., Colavizza, G., Peroni, S. (2026). OC-AND (Version v1.0.0) [Computer software]. Zenodo. \url{https://doi.org/10.5281/zenodo.23055587}

\section*{Author Contributions}
F.C. designed the study, executed the Bridge4AND pipeline, conducted 
experiments, and wrote the manuscript. G.C. and S.P. supervised the 
work, contributed to methodological design and interpretation of results, 
and revised the manuscript. All authors reviewed and approved the 
final version.

\section*{Competing Interests}
The authors declare no competing interests.

\section*{Acknowledgements}
We acknowledge WhoIsWho, OpenCitations infrastructure and Crossref API 
for providing open metadata access essential to this work.

\section*{Funding}
This work was funded by the European Union's Horizon Europe Research and Innovation programme under Grant Agreement No. 101187940 (LUMEN project).

\section*{Declaration of Use of Generative AI}
No generative AI systems were used to generate or analyse data, or draft the scientific content of this manuscript. AI-assisted tools were used only for spell-checking and minor grammar suggestions on the English text. Generative AI–based coding assistants were used during the development of Bridge4AND to support code generation and, especially, review; all AI-suggested code was reviewed, tested, and validated by the authors.

\bibliographystyle{unsrtnat}

\end{document}